\documentclass[journal=aamick,manuscript=article]{achemso}

\usepackage[version=3]{mhchem} 

\newcommand{\spt}{sp$^2$}
\newcommand{\dia}{sp$^3$}
\newcommand{\meth}{CH$_4$}
\newcommand{\hydro}{H$_2$}
\newcommand{\hplus}{H$^+$}

\newcommand{\degC}{$^\circ$C}
\newcommand{\epscomp}{\varepsilon_{\textrm{r}}}
\newcommand{\epsr}{\varepsilon_{\textrm{r}}'}
\newcommand{\epsd}{\varepsilon_{\textrm{r}}''}
\newcommand{\epss}{\varepsilon_{\textrm{s}}}
\newcommand{\epsi}{\varepsilon_{\infty}}

\makeatletter
\def\textless{\afterassignment\textless@\let\next= }
\def\textless@#1#{\@nameuse{textless@#1}}
\def\textless@sub#1#2/sub#3{%
  \ensuremath{_{\let\textless\relax#2}}%
  \egroup 
}
\def\textless@sup#1#2/sup#3{%
  \ensuremath{^{\let\textless\relax#2}}%
  \egroup 
}
\makeatother

\author{Jerome A. Cuenca}
\affiliation[Cardiff University]{Cardiff School of Physics and Astronomy, Cardiff, Wales, CF24 3AA, UK}
\email{cuencaj@cardiff.ac.uk}
\author{Soumen Mandal}
\affiliation[Cardiff University]{Cardiff School of Physics and Astronomy, Cardiff, Wales, CF24 3AA, UK}
\author{Malcolm Snowball}
\affiliation{Ultra Biotecs Limited, Derby, DE24 9FU, UK}
\author{Adrian Porch}
\affiliation[Cardiff University]{Cardiff School of Engineering, Cardiff, Wales, CF24 3AA, UK}
\author{Oliver A. Williams}
\affiliation[Cardiff University]{Cardiff School of Physics and Astronomy, Cardiff, Wales, CF24 3AA, UK}

\title[Dielectric spectroscopy of hydrogenated hexagonal boron nitride ceramics]
  {Dielectric spectroscopy of hydrogenated hexagonal boron nitride ceramics}

\keywords{boron nitride, dielectric spectroscopy, complex permittivity, hydrogenation}

\begin{document}

\begin{tocentry}

\includegraphics[width=\textwidth]{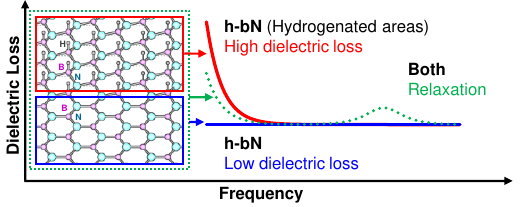}

\end{tocentry}

\begin{abstract}
   Hexagonal boron nitride (h-BN) is a critical material for 2D electronic devices for graphene and has attracted a considerable amount of attention owing to its structural similarity and semiconducting property. However, modifying its wide-band gap is a challenge.  Hydrogenation is a potential method of altering the electrical properties, although is seldom experimentally measured. Here, the complex permittivity of h-BN after various hydrogen treatments have been investigated. For untreated h-BN, a frequency independent dielectric constant was measured ($\sim4.2 \pm0.2$) and an immeasurably low dielectric loss, demonstrating the ideal dielectric nature of h-BN across the $10^3$ to $10^{10}$ Hz range. However, after atomic H-plasma treatment in a microwave chemical vapour deposition (CVD) reactor, the complex permittivity amplifies dramatically, introducing dielectric dispersion through Debye-type dielectric relaxations ($\epss\approx20\pm2$, $\epsi\approx4.2\pm0.2$) and a percolating long range conductivity ($\sim0.32$ mS/m). Annealing in molecular hydrogen at similar CVD temperatures showed minimal effect. Raman spectroscopy also detected minimal change in all samples, implying the increase is not due to other phases. This leads to the experimental conclusion that hydrogenation, through atomic H-plasma treatment, results in a moderate increase in room temperature electrical conductivity, an associated finite dielectric loss factor. The potential as a tunable wide-band gap semiconductor is highlighted however for insulating dielectric substrate applications, microwave CVD may destroy these desirable properties.
\end{abstract}

\section{Introduction}
The dielectric properties of hexagonal boron nitride (h-BN) have received recent interest owing to its two-dimensional (2D) structural similarity to graphene, albeit electrically insulating. Since graphene is incredibly sensitive to the environment, dielectric encapsulation using h-BN is a possible stabilisation method\cite{Siskins2019,Barnard2016,Dauber2015}. h-BN has a small lattice mismatch with graphene of approximately 1.5 to 1.7\%\cite{Wang2017,Kumar2016}, a high band gap of 5.97 eV\cite{Wang2017}, high dielectric strength\cite{Ji2016} and a low dielectric constant (ranging from 2 to 5)\cite{Kim2012a,Chen2007}. There are also contrasting dielectric applications of h-BN, including microwave absorbing composites using h-BN as an insulating inclusion, as well as potential biomedical applications\cite{Nose2006,Pang2018,Zhou2012a,Merlo2018}. As a dielectric substrate for graphene, h-BN is commonly produced using thin film sputtering or chemical vapour deposition (CVD). The resultant material may also include unwanted BN phases including amorphous and turbostratic boron nitride (a-BN and t-BN). Recent progress has been made through sputtering onto nanocrystalline diamond substrates, with a lower density of a-BN and t-BN, although at slow deposition rates of hundreds of nanometres per hour\cite{Hoang2017}. The hot pressed method can produce thick h-BN ceramics using boric acid binders, with pores leading to an increased hygroscopicity.  From an electronic device perspective, thicker dielectrics of h-BN are favourable to contain stray electromagnetic fields.

Graphene deposition through CVD involves exposing the substrate to a carrier gas at high temperatures ($>$800 \degC{}). Successful deposition has been demonstrated using microwave plasma CVD in hydrogen (\hydro{}) on Cu substrates\cite{Woehrl2014,Fang2016}. h-BN substrates can survive these high temperatures (stability up towards $\sim$1000 \degC{})\cite{Kostoglou2015}. It would be assumed that the favourable low dielectric constant of the h-BN is unchanged, however, like graphene, \spt{} bonding is prone to etching. Reactive \hplus{} ions are known to etch \spt{} in BN\cite{Konyashin1999}, partially transform \spt{} into \dia{} through hydrogen termination of dangling bonds\cite{Khvostov2000} and diffuse through and distort h-BN layers \cite{He2019}. A change in the polarisability is expected, especially with hydrogenation, though this hypothesis is rarely tested. Also, the  conventional tube furnace CVD methods, where graphene is deposited on h-BN in \hydro{} ($>$800 \degC{}), may also affect the dielectric properties\cite{Son2011,Kim2013,Arjmandi-Tash2018}.

Dielectric spectroscopy of h-BN is an important research area for low dielectric constant graphene substrates and while there is much in the literature on its \textit{calculation}, experimental measurements are uncommon, let alone studies of hydrogenation. Notable works in the radio frequency (RF) domain by Kim et al. demonstrate the favourably low dielectric constant\cite{Kim2012a}, while Shi et al. show layer stacking effects of BN \cite{Shi2014}. Previous dielectric measurements have been measured in the RF or optical range, though rarely at microwave frequencies\cite{Madelung2001}.

This work shows the effect of atomic and molecular hydrogen treatment (plasma and annealing) on the broadband ($10^3$ to $10^{10}$ Hz) dielectric properties of h-BN, achieved using well-established non-destructive dielectric spectroscopy methods, including parallel plate capacitor (PPC), broadband coaxial probe (BCP) and microwave cavity perturbation (MCP)\cite{Cuenca2015a,Cuenca2019}.

\section{Theory}
\subsection{Complex permittivity}
\label{sec:theory}
The complex permittivity is defined as the ability of a material to polarise in an electric field with the imaginary part associated with the time-harmonic polarisation loss:
\begin{equation}
	\epscomp = \epsr(\omega)-j\epsd(\omega)
\end{equation}
where $\epsr(\omega)$ is the frequency dependent dielectric constant, $ \epsd(\omega)$ is the dielectric loss factor, $\omega$ is angular frequency in rad/s or $2\pi f$ and $f$ is frequency in Hz. Dielectric polarisation mechanisms in solids arise from free charge conductivity (Drude model), space charge polarisation in inhomogeneous conducting mixtures or dipolar relaxation (Debye/Havrilliak-Negami models) and bound electronic polarisation (Lorentz model)\cite{Ding2019}. In the case of non-polar materials, dipolar relaxation is non-existent and so too are Lorentz based frequency dependent contributions from electronic relaxation when measuring at much lower than terahertz frequencies. For percolating free charge conductivity or long range conductivity, the dielectric loss factor follows\cite{Eichelbaum2012,Krupka2006}:
\begin{equation}
	\varepsilon_{\textrm{r, Drude}}''(f) \approx \frac{\sigma}{\omega\varepsilon_0}
	\label{eq:cond}
\end{equation}
where $\varepsilon_0$ is the permittivity of free space and $\sigma$ is the free charge conductivity in S/m. A characteristic $\omega^{-1}$ dependence in the dielectric loss implies this mechanism. For space charge polarisation, both real and imaginary parts are modelled by Debye/Havrilliak-Negami type relaxations\cite{Ding2019}:
\begin{equation}
	\varepsilon_{\textrm{r, HN}}\approx\epsi +
	\left( \frac{\epss-\epsi}{\left[1+\left(j\omega\tau\right)^\alpha\right]^\beta}\right)
	\label{eq:debye}
\end{equation}
where $\epss$ and $\epsi$ are the low and high frequency dielectric constants, respectively, $\tau$ is the time constant of the hopping process and $\alpha$ and $\beta$ are fitting exponents as empirical values rarely follow the ideal Debye model (where $\alpha=\beta=1$). The result is a decrease in the real part and a loss peak at the relaxation frequency of $1/\tau$.


\section{Experimental}
\label{sec:method}
The complex permittivity of most materials can be measured, broadly speaking, with two approaches: broadband transmission/reflection and resonant techniques, whereby the former allows the observation of dispersion from frequency dependent polarisation mechanisms while the latter methods offer high narrowband sensitivity.

\subsection{Parallel plate method (PPC)}
The parallel plate method is the simplest non-resonant approach to obtaining the complex permittivity, whereby the material is contacted using two metallic electrodes and impedance is measured. The complex permittivity is related through\cite{Liu2007}:
\begin{equation}
	C - j\frac{G}{\omega} = k_{\textrm{eff}} \varepsilon_{\textrm{r}}
		\label{eq:cap}
\end{equation}
where $C$ is the capacitance, $G$ is the conductance, $k_{\textrm{eff}}$ is a geometrical constant found from an air-measurement or of a known calibration sample, $\varepsilon_0$ is the permittivity of free space, $t$ is the thickness of the sample material. The first term of \ref{eq:cap} can be obtained by first measuring the fixture in free space. The sample must cover the entirety of the electrode, whereby the out-of-plane dielectric value is measured for flat samples.

\begin{figure}[t!]
  \includegraphics[width=0.7\textwidth]{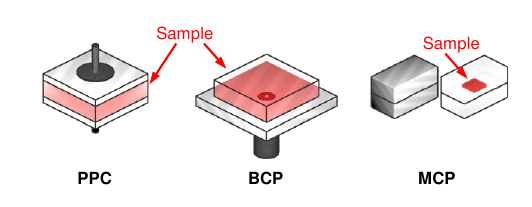}
\caption{Graphical representation of the methods used in this study}\label{fig:methods}
\end{figure}

\subsection{Broadband coaxial probe (BCP)}
The open ended coaxial probe is a reflection based method which uses a similar capacitance perturbing approach in that a coaxial cable is terminated into free space or a dielectric sample. The radial transverse electromagnetic  fields which exist within the cable are suddenly interrupted by a free space section, giving rise to a predominantly parallel field with some perpendicular fields into the sample. Assuming that the sample occupies infinite space at the end of the probe, the complex permittivity can be obtained using the reflection coefficient. This assumption is fulfilled by having a sample several hundreds of microns thick for a millimetre small coaxial aperture. The impedance has been measured here through scattering parameters and the extraction of complex permittivity is approximated as follows\cite{Cuenca2015a}:

\begin{equation}
\varepsilon_{\textrm{r}} \approx \frac{1}{j\omega\varepsilon_0C_0Z_0} \left(\frac{1-\Gamma_{\textrm{L}}/\Gamma_{\textrm{a}}}{1+\Gamma_{\textrm{L}}/\Gamma_{\textrm{a}}}\right) +1
\label{eq:bcp}
\end{equation}
where $\Gamma_{\textrm{L}}$ and $\Gamma_{\textrm{a}}$ are the air and sample terminated complex reflection coefficients, respectively, $Z_0$ is the characteristic impedance taken as 50 $\Omega$ and $C_0$ is the probe capacitance which may be obtained by measuring a material of known dielectric constant. In this study, we have calibrated the BCP to a polytetrafluoroethylene (PTFE) standard.

\subsection{Microwave cavity perturbation (MCP)}
The microwave cavity perturbation method is a resonant technique whereby the sample is placed within the electric ($E$) or magnetic ($H$) fields of a microwave resonator and the presence of the sample within the field alters the resonant frequency of the system. Differences in the unperturbed and perturbed response are attributed to the dielectric or magnetic properties, depending on the volume perturbation within the electric ($E$) or magnetic ($H$) field as given by the following:

\begin{equation}
-\frac{\Delta\omega}{\omega_{\textrm{0}}} \approx \frac{\varepsilon_{\textrm{r}}-1}{1+N(\varepsilon_{\textrm{r}}-1)} \frac{V_{\textrm{s}}}{V_{\textrm{m}}}
\label{eq:depol}
\end{equation}
where $\Delta\omega/\omega_0$ is the fractional change in complex resonance caused by an E-field sample perturbation, $V_{\textrm{s}}$ and $V_{\textrm{m}}$ denote the sample volume and mode volume of the cavity, respectively, $\varepsilon_{\textrm{r}}$ is the complex relative permittivity and $N$ is the geometric sample depolarising factor which is positive and less than or equal to unity\cite{Cuenca2019}. For low permittivity samples placed in minimal depolarising geometry, $N\approx0$ but in other cases, $N$ may be obtained analytically, through finite element modelling, or through measurement of a known sample \cite{Cuenca2019,Cuenca2015}.

The samples used were commercially available hot pressed BN ceramics used in previous studies\cite{Mandal2019b}. The as received h-BN substrates have dimensions of $\sim0.5\times10\times10$ mm. Hydrogen (H) plasma treatment of as received h-BN was carried out using a Seki Technotron AX6500 microwave CVD reactor (4 kW at 50 Torr, 500 sccm of \hydro{} for 1 hour at $\sim800$\degC{}). The substrate temperature was monitored using a Williamson dual wavelength pyrometer. Annealing was carried out using a furnace at $\sim800$ \degC{} in vacuum or H$_2$ ambient (flow rate of 100 sccm, 7.5 Torr). Samples were also dipped in deionised water before annealing to encourage hydrogen uptake.

Dielectric measurements have been carried out from 1 kHz and 10 GHz. The PPC method has been conducted from 1 kHz to 1 MHz (Keysight E4990A impedance analyser and 16451B). For moderately conductive samples, electrode polarisation (EP) introduces an additional inverse power law artefact in PPC ($\propto f^{-\gamma}$)\cite{Colosi2013}. This has been accounted for by assuming a static low frequency dielectric constant and extrapolating the BCP $\epsr$ value, revealing the long range conductivity component. The BCP method has been conducted from 10 MHz to 10 GHz (Keysight N5232A vector network analyser) with an estimated penetration depth of $<$0.2 mm\cite{Cuenca2015a,Porch2012}. MCP has been conducted with an Al rectangular cavity at 2.5, 4.6 and 5.5 GHz\cite{Cuenca2019}. An additional WG14 waveguide cavity was used at 4.5, 5.6, 7.4 to 9.6 GHz to corroborate the Al cavity. For the unperturbed response, an acetate sheet was used to suspend the sample in the centre of the cavity. Prior to sample measurement, a $1.5\times10\times10$ mm$^3$ piece of PTFE was used as a calibration standard. The well-known non-dispersive dielectric nature of PTFE allowed the different methods to be lined up across the frequency range with the PPC value as reference.
\section{Results and discussion}
\label{sec:result}

\subsection{Complex permittivity}
\begin{figure}[t!]
  \includegraphics[width=0.7\textwidth]{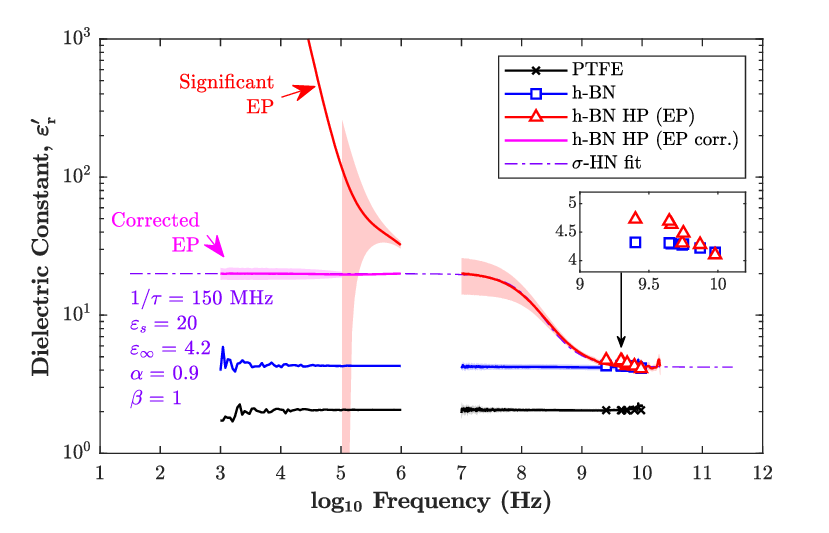}
  \includegraphics[width=0.7\textwidth]{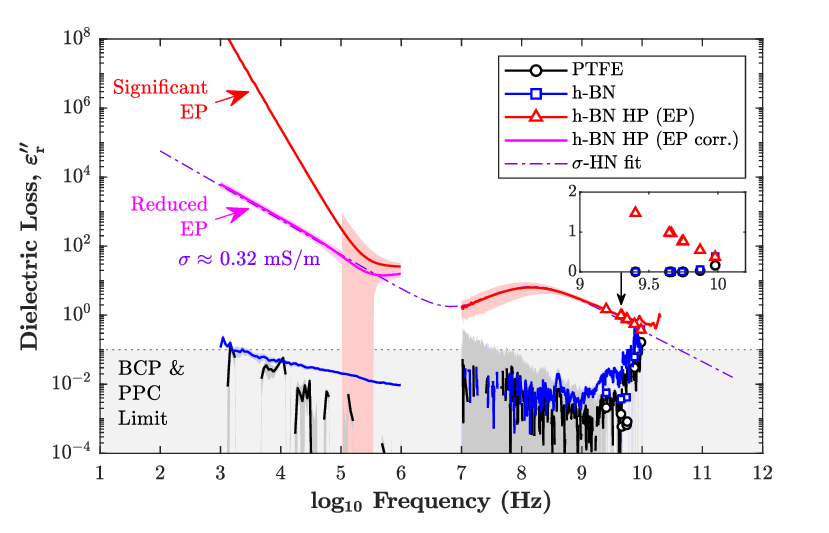}
\caption{Complex permittivity of PTFE, h-BN, and H plasma treated sample (HP) obtained using PPC, BCP and MCP with the $\sigma$-HN model from (\ref{eq:cond}) and (\ref{eq:debye}). EP correction provides an estimate of free charge conductivity. Shaded regions show the uncertainty: the standard deviation of 6 sample measurements and instrumental limitation.}\label{fig:eps}
\end{figure}

\begin{figure}[t!]
  \includegraphics[width=0.7\textwidth]{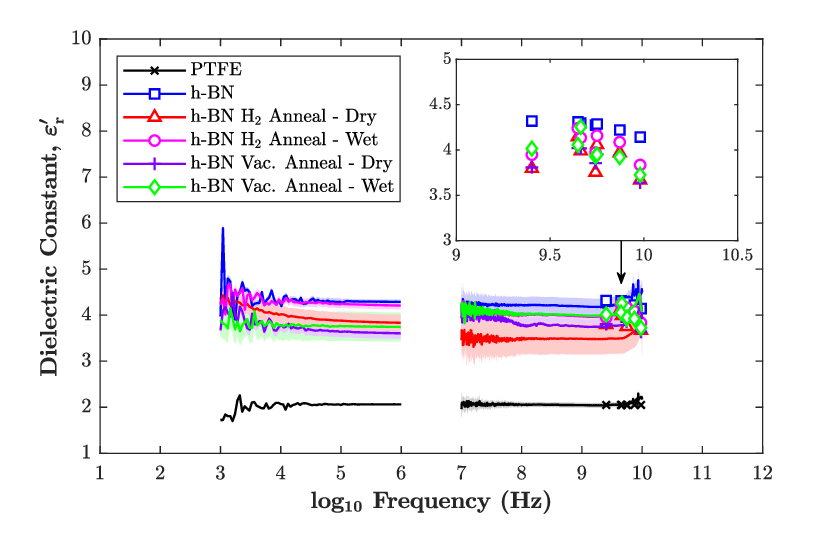}
  \includegraphics[width=0.7\textwidth]{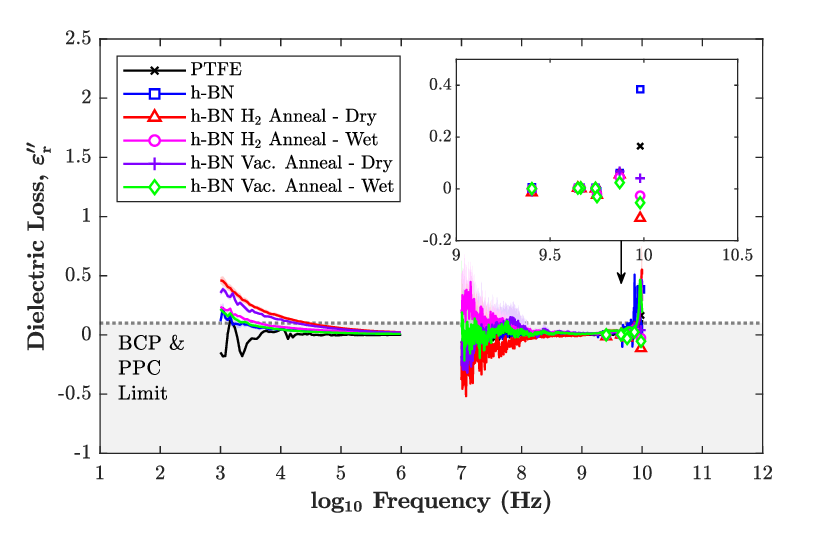}
\caption{Complex permittivity of PTFE standard, h-BN and annealed h-BN samples obtained using PPC, BCP and MCP. The shaded regions mark the uncertainty: the standard error of 6 sample measurements and instrumental limitation.}\label{fig:anneals}
\end{figure}

The measured complex permittivity of the PTFE, untreated, plasma treated and annealed h-BN samples are given in Figures \ref{fig:eps} and \ref{fig:anneals} and Tables \ref{tab:eps1} and \ref{tab:eps2}. The PTFE standard gave a nominal value of 2.05 across the low kilohertz to megahertz frequency range. Since PTFE is assumed non-dispersive, therefore, this value is extrapolated to higher frequencies, allowing the PPC, BCP and MCP methods to be compared for the h-BN samples. It is also shown that there are considerable systematic variations from 1 to 10 kHz which are systematic errors. From Table \ref{tab:eps2} the WG14 MCP measurements showed much higher losses with increasing frequency, particularly at 9.6 GHz. Since it is well-known that PTFE has negligible loss at gigahertz frequencies, the additional loss is likely a systematic error associated with the strong coupling to the resonator. 

The untreated h-BN gave a dielectric constant of 4.3$\pm0.1$ in the kilohertz to megahertz range, in close agreement with the analytically calculated out-of-plane value\cite{Laturia2018}. Even though h-BN is 2D, the material is a compressed ceramic and so an anisotropic value was not obtained, rather an averaged isotropic value. With increasing frequency, there was minimal change with a value of 4.2$\pm0.2$ obtained using the BCP method. In the gigahertz range, MCP corroborated the BCP values as shown in Table \ref{tab:eps1}. As with the PTFE sample, the dielectric loss was immeasurable in the megahertz to gigahertz range, with the same artefact of the slight increase at 9.6 GHz. The negligible loss and low dielectric constant demonstrates the favourable insulating dielectric properties of h-BN.

After H-plasma treatment, the complex permittivity increased dramatically, with dispersive features. Starting with the MCP values at gigahertz frequencies, the dielectric constant increased from the untreated to sample $4.726\pm 0.006$ at 2.5 GHz though remained at a similar value to the untreated of $4.101\pm 0.001$ at 9.6 GHz. The dielectric loss, however, increased substantially from immeasurable to $1.48\pm 0.01$ at 2.5 GHz and decreasing to $0.38\pm 0.01$ at 9.6 GHz, characteristic of a relaxation process occurring at lower megahertz frequencies. The BCP data confirmed this relaxation mechanism, with an approximate Havriliak-Negami model ($\varepsilon_s = 20$, $\varepsilon_{\infty} = 4.2$, $\tau^{-1} = 150$ MHz, $\alpha = 0.9$ and $\beta = 1$). Finally, towards lower frequencies, the conductivity was large enough to induce EP in the PPC method. The corrected data, extrapolated from the BCP dielectric constant, revealed a percolating free charge conduction mechanism of approximately 0.32 mS/m. This value was corroborated with contacting multi-meter measurements of 0.1 to 0.5 mS/m. All other samples could not be measured due to their high resistivity. To verify if the dielectric enhancement was a result of the elevated temperatures in a \hydro{} atmosphere, annealing in both vacuum and \hydro{} was performed. Figure \ref{fig:anneals} shows the results, with a small decrease in the overall dielectric constant. A small increase in the low frequency loss was measured upon dry annealing and the attempt to encourage \hydro{} uptake through wetting gave minimal change.


\begin{table*}[t]
 \caption{Tabulated dielectric constant of PTFE and h-BN samples}
 \label{tab:eps1}\footnotesize
\begin{tabular*}{\textwidth}{@{\extracolsep{\fill}}llllllllll}
\hline
  	Method*		&	PPC		 &	BCP		&	\multicolumn{3}{l}{MCP (Al Cavity)}  &	\multicolumn{4}{l}{MCP (WG14 Cavity)} \\
  	$f$ 	(GHz)		&	$10^{\textrm{-6 to -3}}$		&	$10^{\textrm{-2 to 1}}$ 		&	2.5 & 4.6 & 5.5 & 4.5 & 5.6 & 7.4& 9.6\\
\hline
PTFE	&	2.05	&	2.05							
	&	2.05	&	2.05	&	2.05					
	&	2.05	&	2.05	&	2.05	&	2.05		\\
											
h-BN	&	4.3	&	4.2					
	&	4.320	&	4.298	&	4.275				
	&	4.309	&	4.287	&	4.221	&	4.143		\\

h-BN (H-Plasma)	&	20	&	20 to 4.3							
	&	4.726	&	4.636	&	4.309					
	&	4.691	&	4.482	&	4.282	&	4.101		\\

h-BN (\hydro{} Anneal, Dry)	&	3.9	&	3.6							
	&	3.796	&	3.991	&	3.754					
	&	4.146	&	4.060	&	3.969	&	3.669	\\

h-BN (\hydro{} Anneal, Wet)	&	4.3	&	4.2							
	&	3.948	&	4.135	&	3.972					
	&	4.240	&	4.158	&	4.087	&	3.835	\\
	
h-BN (Vac. Anneal, Dry)	&	3.7	&	4.5							
	&	3.807	&	4.018	&	3.855					
	&	4.021	&	3.951	&	3.940	&	3.640	\\

h-BN (Vac. Anneal, Wet)	&	3.8	&	4.2							
	&	4.016	&	4.254	&	3.931					
	&	4.056	&	3.949	&	3.922	&	3.725	\\
\hline
\multicolumn{10}{l}{\resizebox{\columnwidth}{!}{*Maximum standard deviation of 6 samples of approximately $\pm0.1$, $\pm0.2$ and $\pm6\times10^{-3}$ for PPC, BCP and MCP, respectively.}}
  \end{tabular*}
\end{table*}

\begin{table*}[t]
 \centering
 \caption{Tabulated dielectric of loss PTFE and h-BN samples}
  \label{tab:eps2}\footnotesize
\begin{tabular*}{\textwidth}{@{\extracolsep{\fill}}llllllllll}
\hline
  	Method*		&	PPC		 &	BCP		&	\multicolumn{3}{l}{MCP (Al Cavity)}  &	\multicolumn{4}{l}{MCP (WG14 Cavity)} \\
  	$f$ 	(GHz)		&	$10^{\textrm{-6 to -3}}$		&	$10^{\textrm{-2 to 1}}$ 		&	2.5 & 4.6 & 5.5 & 4.5 & 5.6 & 7.4& 9.6\\
\hline
PTFE	&$	0	$&$	0	$						
	&$	0.002	$&$	0	$&$	0	$				
	&$	0.006	$&$	0	$&$	0.03	$&$	0.2		$\\
											
h-BN	&$	0	$&$	0		$					
	&$	0.006	$&$	0.004	$&$	0.004	$				
	&$	0.01	$&$	0	$&$	0.06	$&$	0.4		$\\

h-BN (H-Plasma)	&   $0.00032/\omega\varepsilon_0$ &$	6.4$ to 0							
	&$	1.48	$&$	0.97	$&$	0.76	$			
	&$	3.97	$&$	0.77	$&$	0.55	$&$	0.38		$\\

h-BN (Vac. Anneal, Dry)	&$	0.5 $ to 0&$	0			$				
	&$	0	$&$	0	$&$	0			$		
	&$	0.009	$&$	0	$&$	0.07	$&$	0.04		$\\

h-BN (Vac. Anneal, Wet)	&$	0.2	$ to 0&$	0		$					
	&$	0	$&$	0.004	$&$	0.002$					
	&$	0.01	$&$	0	$&$	0.02	$ &$	0		$\\

h-BN (\hydro{} Anneal, Dry)	&$	0.4	$ to 0&$	0			$				
	&$	0	$&$	0	$&$	0			$		
	&$	0.01	$&$	0	$&$	0.06	$&$	0 	$\\

h-BN (\hydro{} Anneal, Wet)	&$	0.2	$ to 0&$	0			$				
	&$	0	$&$	0.003	$&$	0.002	$				
	&$	0.02	$&$	0	$&$	0.05	$&$	0	$\\

\hline
\multicolumn{10}{l}{\resizebox{\columnwidth}{!}{*Maximum standard deviation of 6 samples of approximately $\pm0.1$, $\pm0.2$ and $\pm6\times10^{-3}$ for PPC, BCP and MCP, respectively.}}
  \end{tabular*}
\end{table*}

\subsection{Raman Spectroscopy}
\begin{figure}[t!]
  \includegraphics[width=0.7\textwidth]{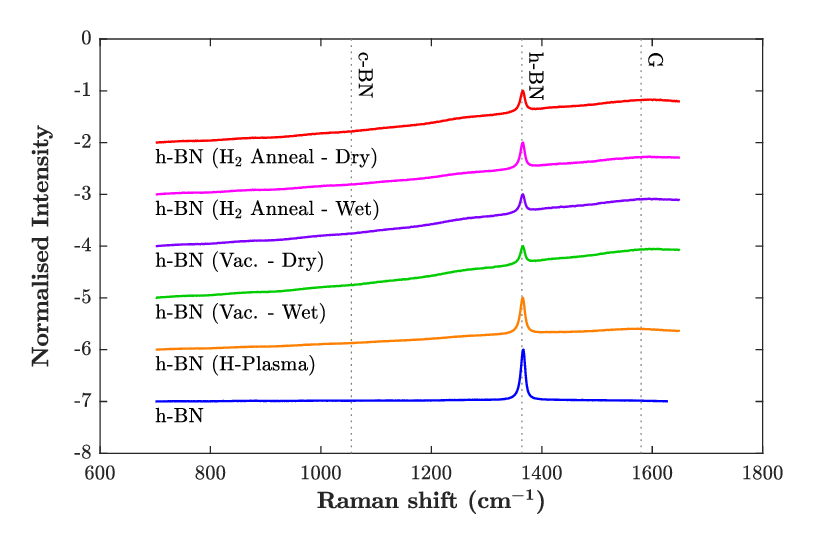}
\caption{Raman Spectroscopy (top) of hydrogen treated h-BN samples. All data has been normalised to the h-BN peak. No significant variations in composition are noticed although a small G-band is found in all treated samples.}\label{fig:raman}
\end{figure}

\begin{figure}[t!]
     \centering
    \begin{tabular}{ccc}
        \includegraphics[width=0.3\textwidth]{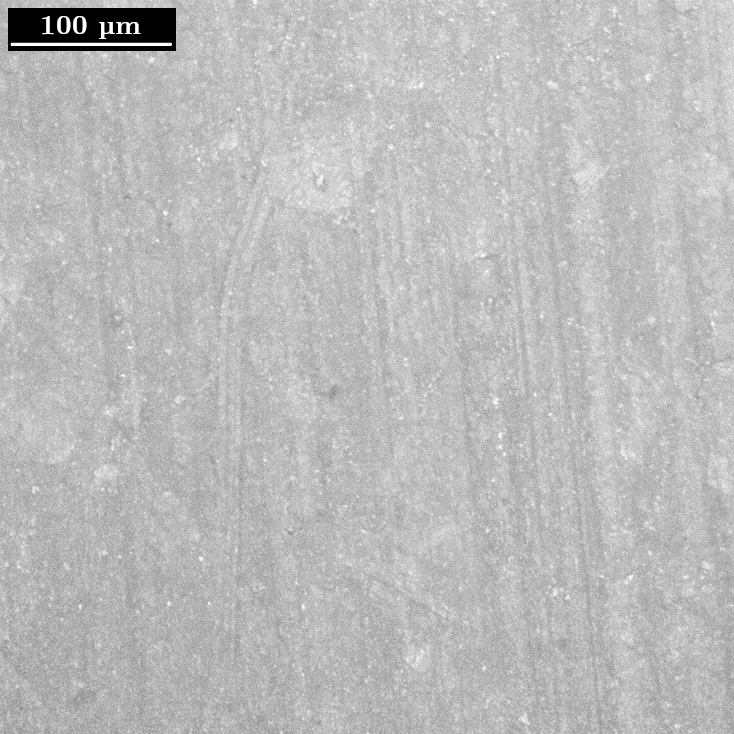} &
        \includegraphics[width=0.3\textwidth]{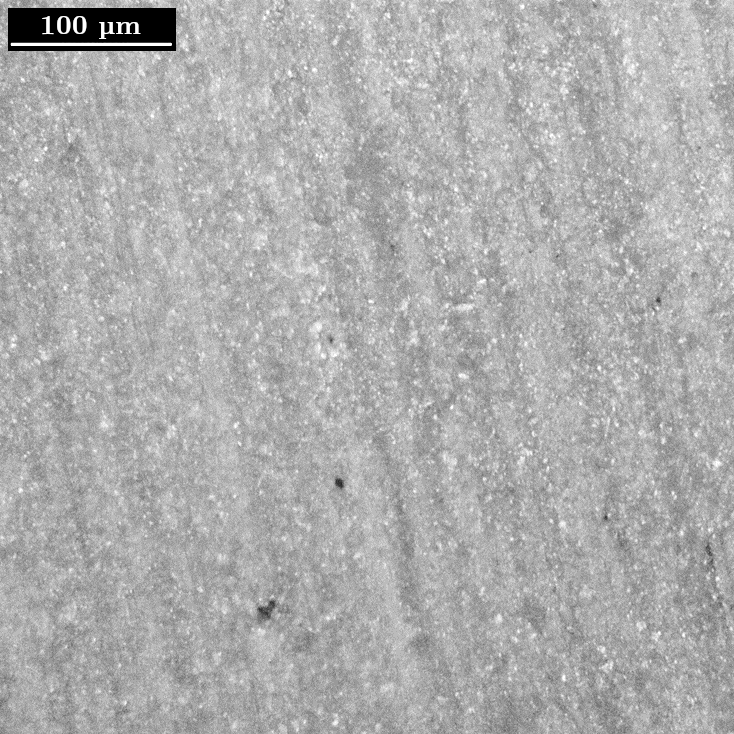} &
        \includegraphics[width=0.3\textwidth]{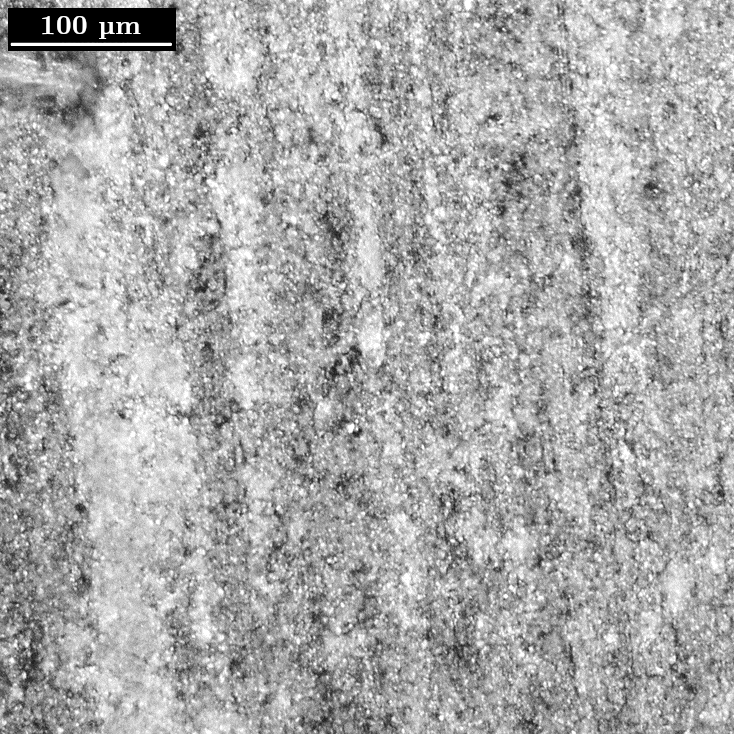} \\
        (a) & (b) & (c)
    \end{tabular}
    \caption{Microscope images (20$\times$ magnification) of (a) pristine h-BN , (b) H$_2$ dry annealed  and (c) H-Plasma treated samples.}\label{fig:mic}
\end{figure}

Raman data of the samples are shown in Fig. \ref{fig:raman}, revealing that the predominant phase in the pristine BN ceramic is hexagonal (1364 cm$^{-1}$), with almost no detectable cubic phase (1055 cm$^{-1}$)\cite{Mandal2019b}. After H-plasma treatment, there is minimal change in the spectra, although signs of background fluorescence. Curiously, a very small band is apparent at approximately 1580 cm$^{-1}$ which is attributed to the G-Band of \spt{} carbon. This is most likely associated with trace carbon contaminants that have been incorporated into the ceramics from the  sample holder. Similarly for the \hydro{} and vacuum annealed samples, the h-BN peak is apparent with the same small G-Band. However, the origin of this band is most likely associated with contaminants from the graphite crucible during annealing. Equally, a background fluorescence was also measured. This is an interesting result and fully demonstrates that the potential cause of the increase in the dielectric property is less likely to be related to \spt{} carbon contamination from the microwave CVD system, even though \spt{} carbon is known to increase the complex permittivity of various materials\cite{Cuenca2019,Lv2016}. Microscope images of the samples in Fig. \ref{fig:mic} demonstrate that there was minimal visual differences between the pristine and the annealed sample; similar images were obtained for the wet and vacuum annealed samples. The plasma treated sample, however, showed significant discoloration.

\label{sec:discussion}
The untreated h-BN results are similar to the calculated values reported by several authors\cite{Laturia2018,Kim2012a,Kumar2016}. For h-BN, the 2D structure inherently implies anisotropy, to which the anticipated bulk complex permittivity in-plane ($\varepsilon_{\textrm{r,}\parallel}$) is dissimilar to the out-of-plane value ($\varepsilon_{\textrm{r,}\perp}$); calculations demonstrate that electronic polarisation predominantly contributes with $\varepsilon_{\textrm{r,}\parallel}>\varepsilon_{\textrm{r,}\perp}$\cite{Laturia2018}. Even though the field orientations vary amongst PPC, BCP and MCP, anisotropy is not measured implying that the macroscopic dielectric properties of hot pressed h-BN ceramics is isotropic.

The most pertinent finds in this study are (i) the dielectric properties of h-BN when annealed in \hydro{} at $\sim800$ \degC{} are minimally changed, (ii) when exposed to atomic H, a huge amplification occurs with an associated relaxation in the megahertz range in addition to (iii) a percolating long range conduction mechanism.

In the case of (i), the temperature stability of h-BN has been demonstrated and that tube furnace type CVD methods to deposit graphene are seemingly least likely to damage the insulating properties of the h-BN. The peculiar result of (ii) demonstrates that atomic H leads to an enhanced conductivity. The enhanced dielectric property likely stems from a form of Maxwell-Sillars-Wagner type polarisation of an insulating host medium with small conducting regions\cite{Sihvola1999}. In an electric field, the charges migrate and are impeded by the insulating boundaries, resulting in charge build-up and an amplified capacitance effect. At shorter timescales, or higher frequencies, these charges never polarise resulting in relaxation. For the plasma treated h-BN, however, regions of some conducting phase must exist. There are two possibilities for this with the first being due to B or N vacancy formation as the BN is etched in the hydrogen plasma and the full or partial hydrogenation of h-BN.

There are numerous density functional theory (DFT) studies on the vacancy formation and hydrogenation in h-BN sheets which all show that the electronic properties of h-BN are altered\cite{Spath2017,Huang2012,Wang2010}. Huang and Lee demonstrated using a DFT model that introducing vacancies at both B and N sites creates several mid-gap states, although seemingly deep at several eV from the valence band minimum\cite{Huang2012}. This implies that vacancy formation is not the principal cause of the increased room temperature conductivity observed in the current study.

In the argument pertaining to hydrogenation, the exposure of the h-BN ceramic to atomic hydrogen results in hydrogen termination of individual islands of h-BN which are separated by the binder. One theory is that hydrogen termination is known to produce a negative electron affinity (NEA)\cite{Powers1995}, which in turn promotes surface conductivity\cite{He2015}, similar to that in hydrogen terminated diamond\cite{Williams2003,Landstrass1989}. Although, full hydrogenation of a BN sheet is likely not responsible for the enhanced conduction losses as it is also known that while this does indeed decrease the wide band gap, it does not create significant mid gap states to allow room temperature conductivity\cite{Wang2010}. It was, however, calculated that metallic conduction is possible with semi-hydrogenated BN layers, whereby  only the B atoms are bonded to hydrogen\cite{Wang2010}. This form of semi-hydrogenation, as opposed to full hydrogenation and just hydrogen bonding to N atoms, is seemingly stable and gives rise to a metallic BN character, in addition to a distorted and buckled lattice structure. 

\begin{figure}[t!]
     \centering
        \includegraphics[width=0.7\textwidth]{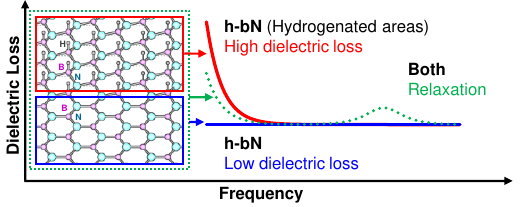}
    \caption{Representative model of low and high dielectric loss regions from pristine and hydrogenated h-BN, respectively. Long range conduction results from percolating hydrogenated h-BN pathways whereas finite regions result in Debye type relaxation.}\label{fig:render}
\end{figure}

Thus, the most plausible explanation behind the increased dielectric properties of the h-BN is that the hydrogen plasma exposure results in locally high dielectric loss, or electrically conducting, islands of semi-hydrogenated h-BN, as is shown in the representation given in Fig. \ref{fig:render}. This also explains why the \hydro{} annealed h-BN yielded minimal change in $\epscomp$, as molecular hydrogen does not hydrogenate the h-BN layer, as this is only achieved with the high energy density plasma. Further precedence to this is that several DFT studies demonstrate that \hydro{} bonding to BN is not stable\cite{Wang2010}. In the Maxwell-Sillars-Wagner model, the characteristic time of the short range charge conduction through one of the islands is $\tau\approx 1/150\textrm{ MHz} \approx 6.7$ ns while longer percolating pathways are also formed resulting in (iii) long range conductivity.

This result demonstrates that microwave plasma CVD of any material onto a h-BN dielectric substrate is ill-advised. As a demonstration, a similar dielectric amplification result was obtained when several microns of CVD diamond was deposited on the h-BN sample. Using the same conditions except with an additional 5\% \meth{} in the gas phase, the amplification was larger ($\epss\approx45$, $\epsi \approx8$, $\tau^{-1}\approx 770$ MHz) in addition to a finite long range conductivity of 8 mS/m (see Supplementary Data). The larger conductivity reported here is due to the additional contributions from the non-diamond carbon impurities and potential leeching of boron into the diamond layer, however, these mechanisms are out of scope of this study. Moving forward with other new and exciting applications of h-BN, however, h-BN is a promising hydrogen storage material to which the detection of changes in dielectric property may infer the concentration of adsorbed atomic hydrogen on B sites. This of course can be easily achieved with the above non-destructive and non-contact methods (MCP in particular)\cite{Lale2018,Hartley2015,Merlo2018}.

\section{Conclusion}
In conclusion, it is demonstrated that the complex permittivity of h-BN has a low dielectric constant and an immeasurably low dielectric loss across the kilohertz to gigahertz frequency range. A dramatic increase in complex permittivity is observed after exposure to atomic hydrogen, resulting in dispersive features. The importance of this work draws attention to the fact that while popular microwave plasma CVD methods are capable of producing graphene on Cu substrates, the desire to move away from metal catalysts and use similar techniques for h-BN must be approached with caution. The annealing studies here demonstrate that adopting tube furnace CVD techniques are a lower risk approach to retaining low dielectric constant h-BN when directly depositing graphene. However, for sensing and other electronic applications, it has been demonstrated that the dielectric properties of h-BN can be tuned through hydrogenation in a microwave plasma.

\begin{acknowledgement}

JAC acknowledges financial support of the Engineering and Physical Sciences Research Council under the program Grant GaN-DaME (EP/P00945X/1). SM and OAW acknowledge financial support of the European Research Council (ERC) Consolidator Grant SUPERNEMS, Project ID: 647471.
\end{acknowledgement}




\bibliography{achemso-demo}

\end{document}